\documentclass[letterpaper,journal]{IEEEtran}

\ifCLASSINFOpdf
\else
\fi
\usepackage[bottom=1.3cm,top=1.3cm,left=1.5cm,right=1.5cm]{geometry}
\usepackage{graphicx}
\usepackage{booktabs}
\usepackage[utf8]{inputenc}
\usepackage{amsmath}
\usepackage{amssymb}
\usepackage{amsthm} 
\usepackage{subfigure}
\usepackage{array}
\usepackage{multirow}
\usepackage{float}
\usepackage{xcolor}
\usepackage{epsfig}

\begin{document}

\title{\huge Optimal Precoding Design for Monostatic ISAC Systems: MSE Lower Bound and DoF Completion}

\author{Yuanhao~Cui,~\IEEEmembership{Member,~IEEE,}
        Fan~Liu,~\IEEEmembership{Member,~IEEE,}
        Weijie Yuan,~\IEEEmembership{Member,~IEEE,}\\
        Junsheng Mu,~\IEEEmembership{Member,~IEEE,}
        Xiaojun Jing,~\IEEEmembership{Member,~IEEE,}
        and~Derrick Wing Kwan Ng,~\IEEEmembership{Fellow,~IEEE}
        \vspace{-2em}
\thanks{Y. Cui, J. Mu, and X. Jing are with the School of Information and Communication Engineering, Beijing University of Posts and Communications (BUPT), Beijing 100876, China. (email:\{cuiyuanhao, mujs, jxiaojun\}@bupt.edu.cn).}

\thanks{F. Liu and W. Yuan is with the Department of Electrical and Electronic Engineering,
Southern University of Science and Technology, Shenzhen 518055, China.
(e-mail: \{liuf6,yuanwj\}@sustech.edu.cn).}

\thanks{D. W. K. Ng are with the School of Electrical Engineering and Telecommunications, University of New South Wales, Sydney, NSW
2052, Australia. (e-mail: w.k.ng@unsw.edu.au).}
}

\maketitle

\begin{abstract}
In this letter, we study the parameter estimation performance for monostatic downlink integrated sensing and communications (ISAC) systems. In particular, we analyze the mean squared error (MSE) lower bound for target sensing in the downlink ISAC system that reveals the suboptimality in re-using the conventional communication waveform for sensing. To realize a practical dual-functional waveform, we propose a waveform augmentation strategy that imposes an extra signal structure, namely the degrees-of-freedom (DoF) completion method. The proposed approach is capable of improving the parameter estimation performance of the ISAC system and achieving the derived MSE lower bound. To improve the performance of the proposed strategy, we formulate an MSE minimization problem to design the ISAC precoder, subject to the communication users' signal-interference-plus-noise-ratio (SINR) constraints. 
Despite the non-convexity of the waveform design problem, we obtain its globally optimal solution via semi-definite relaxation (SDR) and the proposed constructive method. Simulation results validate the proposed DoF completion technology could achieve the derived MSE lower bound and the effectiveness of the MSE-based ISAC waveform design.
\end{abstract}


\IEEEpeerreviewmaketitle

\section{Introduction}

\IEEEPARstart{B}{enefiting} from the improved spectral-, energy- and hardware efficiency, and the ability to deploy sensing functionality into the current communication networks, the research interest for integrated sensing and communication (ISAC) has arisen in the design of the sixth-generation (6G) systems \cite{liu2021integrated}. Depending on the geographical configurations, current ISAC systems are categorized into three classical configurations \cite{li2022assisting}, i.e., 1) the monostatic deployment that transmits communication signals and then captures the target echoes via the co-located receiver, such as enabling a base station (BS) as a sensor; 2) the bistatic deployment that collects the reflected and scattered echoes from a separated receiver, such as various Wi-Fi sensing applications; 3) the distributed deployment that characterizes a target via signals collected from widely distributed receivers. In particular, the monostatic ISAC systems are appealing for practical implementation as it promises a pilot-free signaling strategy \cite{liu2021integrated}.

To unlock the potential integration gain promised by ISAC signaling \cite{cui2021integrating}, one of the major challenges is to design a fully unified waveform design via jointly considering both the sensing and the communication performances. As such, several ISAC signaling strategies have been proposed to strike a balance between communications and sensing \cite{Liu2018TSP}, e.g., embedding communication data into a sensing waveform \cite{hassanien2016dual} and the direct use of standard-compatible communication waveforms for sensing \cite{sit2011ofdm}. Yet, there are limited discussions regarding the estimation error for sensing functionality in ISAC systems, which is also a fundamental performance metric \cite{Zhang2022CST}.

%Compared to the dedicated sensing or communication systems, ISAC technology experiences integration gain or/and coordination gain, which attracts extensive research attention from both industry and academia.

In this letter, we focus on the downlink monostatic ISAC system where an ISAC BS serves multiple users and simultaneously sensing the surrounding environment. In particular, we analyze the parameter estimation performance of the re-used communication waveform, by formulating a transmit waveform design problem that minimizes the MSE for the target sensing. To this end, a MSE lower bound is derived that paves the way for establishing a bound-achieving waveform augmentation strategy, namely \textit{degrees-of-freedom (DoF) completion}, that facilitates the transmit waveform to achieve the MSE lower bound.
%, taking into account the signal-to-interference-plus-noise ratio (SINR) of communication users. 
The simulation results validate the performance of proposed DoF completion strategy and the effectiveness of the MSE-based ISAC waveform design. We highlight our contributions as following:

\begin{itemize}
    \item A MSE lower bound is derived for the monostatic downlink ISAC system, which indicates that re-using conventional communication waveforms without further processing could not achieve the lower bound even if the signal-interference-plus-noise-ratio (SINR) asymptotically high.
    \item A waveform augmentation strategy, i.e., DoF completion, is systematically introduced. Comparing to other pioneer work \cite{liu2021cram,Huang2020TSP}, we confirm that the proposed strategy can achieve the derived MSE lower bound, and then, apply it to a newly proposed MSE-based ISAC waveform design problem. Finally, we obtain its globally optimal solution via semidefinite relaxation (SDR) by constructively showing that the adopted SDR is tight.
\end{itemize}
\textbf{Notations}: Matrices are denoted by bold uppercase letters, vectors are represented by bold lowercase letters; $\text{Tr}(\cdot)$ and $\text{vec}(\cdot)$ denote the trace and the vectorization operations, $(\cdot)^T$, $(\cdot)^H$, $\text{rank}(\cdot)$, and $(\cdot)^{-1}$ stand for the transpose, the Hermitian transpose, the rank, and the inverse of the input matrices, respectively. $\otimes$, and $\mathbb{E}(\cdot)$ denotes the Kronecker product and the expectation operation, respectively. $\mathbf{I}$ represents the identity matrix.

\begin{figure}[!t]
	\centering 
	\includegraphics[width=0.6\columnwidth]{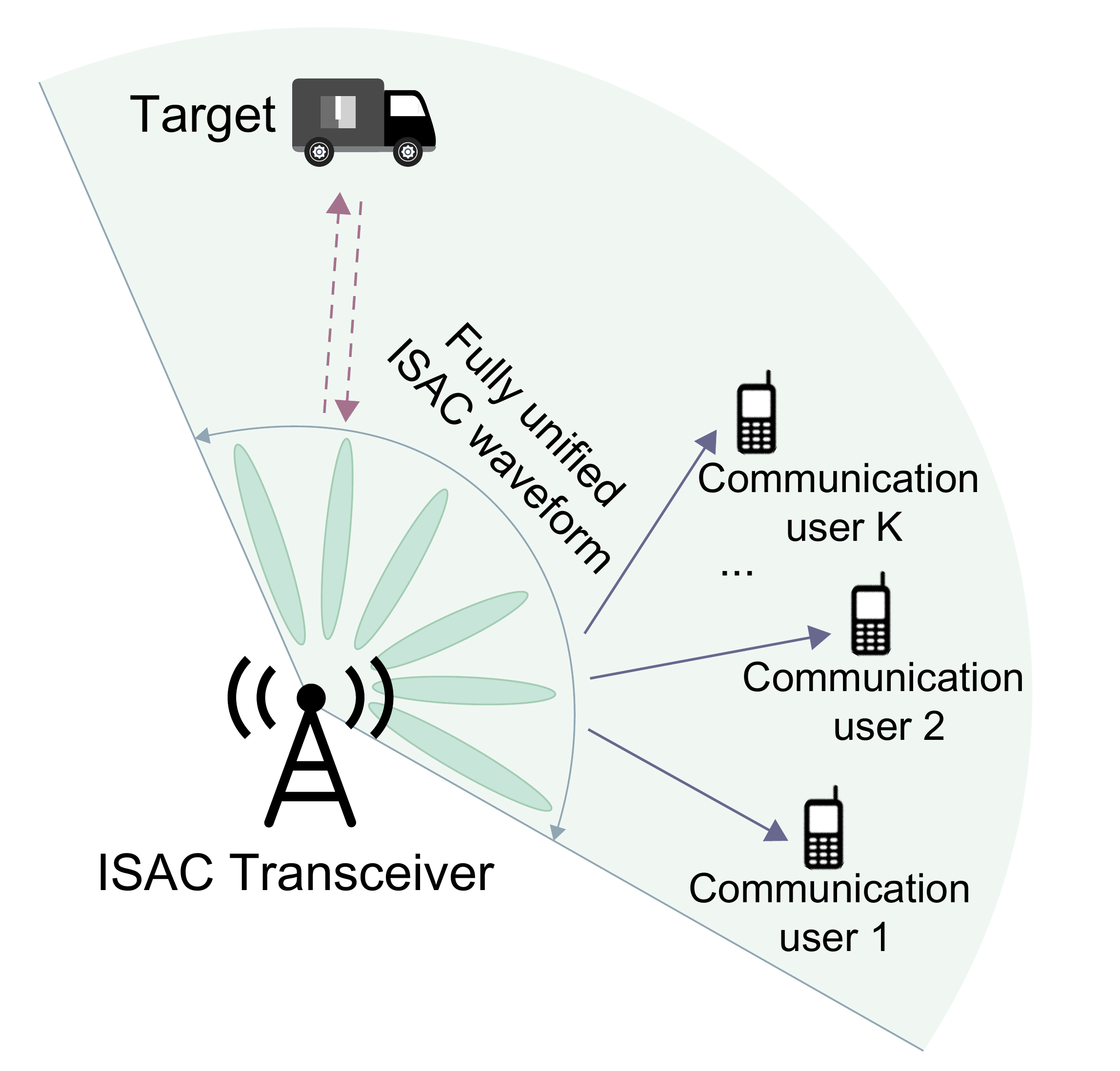}
	\caption{Illustration of the monostatic downlink ISAC system with $K$ communication users and a target.}
	\label{Fig1}
\end{figure}

\section{System Model}
\subsubsection{Communication Model} We consider an integrated sensing and communication BS equipped with $N_T$ transmitting antennas and $N_R$ receiving antennas, as shown in Fig.~\ref{Fig1}. The BS transmits data to $K$ single-antenna communication users and simultaneously extracting environmental information from reflected echoes such that $K \leq N_T$. The channel matrix from the BS to $\mathcal{K} = \{1,\cdots,K\}$ users is represented by $\mathbf{H} \in \mathbb{C}^{K \times N_T} =[\mathbf{h}_1,\cdots,\mathbf{h}_K]^H$, with each $\mathbf{h}_k \in \mathbb{C}^{N_T \times 1}$ denoting the channel vector from the BS to user $k \in \mathcal{K}$. We further assume that the communication channel experiences block fading, in which the data symbol $s_{k,l} \in \mathbb{C}$ transmitted to user $k$ over time slot $l \in \mathcal{L} = \{1,\cdots,L\}$ is precoded by a spatial precoding vector $\mathbf{p}_k \in \mathbb{C}^{K \times 1}$, where $L$ is the channel coherence time interval with $K \leq N_T \leq L$ \cite{Liu2018TSP,Huang2020TSP}. Taking the inter-user interference into account, the received signal at the $k$-th user is given by
\begin{equation}
    y_{k,l} = \mathbf{h}_k^H \mathbf{p}_k s_{k,l} + \sum_{j \in \mathcal{K}/k}\mathbf{h}_k^H \mathbf{p}_j s_{j,l} + n_{k,l}, \forall k \in \mathcal{K}, l \in \mathcal{L},
\end{equation}
where $n_{k,l} \sim \mathcal{CN}(0,\delta_{\text{C}}^2)$ is the additive complex Gaussian white noise at user $k$. Without loss of generality, we assume that the channel knowledge is perfectly known at the BS. Then, the achievable rate at user $k$ can be written as
\begin{equation}
    \mathcal{C}_k(\mathbf{P}) = \log_2(1+\gamma_k(\mathbf{P})),
\end{equation}
where $\mathbf{P} \in \mathbb{C}^{N_T \times K} =[\mathbf{p}_1,\cdots,\mathbf{p}_K]$ denotes the overall precoding matrix at the BS transmitter and $\gamma_k(\mathbf{P})$ is the received SINR at user $k$. By stacking the data symbols transmitted at slot $l$ into a vector $\mathbf{s}_l = [s_{1,l},\cdots, s_{K,l}]^T$, we assume that the transmitted data symbols are independent with unitary power, i.e., $\mathbb{E}(\mathbf{s}_l\mathbf{s}_l^H) = 1$, such that the instantaneous per-user SINR can be given by
\begin{equation}
    \gamma_k(\mathbf{P}) = \frac{|\mathbf{h}_k^H\mathbf{p}_k|^2}{\sum_{j \in \mathcal{K}/k}|\mathbf{h}_k^H \mathbf{p}_j|^2+\delta_{\text{C}}^2}, \forall k \in \mathcal{K}.
\end{equation}
\subsubsection{Sensing Model} The BS re-uses the transmitted waveform for sensing the surrounding environment. The corresponding sensing waveform $\mathbf{X} = \mathbf{P}\mathbf{S}$ is an $N_T \times L$ dimensional matrix that jointly governed by the precoder and transmitted data symbol matrices, where $\mathbf{S} = [\mathbf{s}_1,\cdots,\mathbf{s}_L]$ is a $K \times L$ dimensional data matrix with orthogonal data streams $\frac{1}{L}\mathbf{S}\mathbf{S}^H = \mathbf{I}$. Thus, the sample covariance matrix of $\mathbf{X}$ can be written as
\begin{equation} \label{X2P}
    \mathbf{R}_{\text{X}} = \frac{1}{L}\mathbf{X}\mathbf{X}^H = \frac{1}{L}\mathbf{P}\mathbf{S}\mathbf{S}^H\mathbf{P}^H = \mathbf{P}\mathbf{P}^H =   \mathbf{R}_{\text{P}}.
\end{equation}
Then, the reflected sensing echo $\mathbf{Y}_{\text{S}} \in \mathbb{C}^{N_R \times L}$ received at the BS can be written in a matrix form as
\begin{equation}\label{matrixmodel}
    \mathbf{Y}_{\text{S}} = \mathbf{G}\mathbf{X} + \mathbf{N}_{\text{S}},
\end{equation}
where $\mathbf{G} \in \mathbb{C}^{N_R \times N_T}$ is the target response matrix and $\mathbf{N}_{\text{S}} \in \mathbb{C}^{N_R \times L}$ is the noise matrix. For sensing, we aim for estimating all target responses, i.e., elements of matrix $\mathbf{G}$. For the ease of presentation, we adopt the following operator, $\text{vec}(\mathbf{A}\mathbf{B}) = (\mathbf{B}^T \otimes \mathbf{I}) \text{vec}(\mathbf{A})$, to vectorize \eqref{matrixmodel} as
\begin{equation}\label{vecmodel}
    \mathbf{y}_{\text{S}} = (\mathbf{X}^T \otimes \mathbf{I}) \mathbf{g} + \mathbf{n}_{\text{S}} = \bar{\mathbf{X}}\mathbf{g} + \mathbf{n}_{\text{S}},
\end{equation}
where $\mathbf{g} = \text{vec}(\mathbf{G})$, $\mathbf{n}_{\text{S}} = \text{vec}(\mathbf{N}_{\text{S}})$, and $\bar{\mathbf{X}} = \mathbf{X}^T \otimes \mathbf{I}$ denotes the $N_TN_R$ dimensional vectorized target response matrix, noise vector, and the equivalent sensing waveform matrix, respectively. Typically, the target information is unknown to the sensing system. Hence, we assume that $\mathbf{g}$ is a complex Gaussian random vector $\mathbf{g} \sim \mathcal{CN}(\mathbf{0},\mathbf{R}_\text{g})$, and is independent of the noise vector $\mathbf{n}_{\text{S}} \sim \mathcal{CN}(\mathbf{0},\delta_{\text{S}}^2\mathbf{I})$. 

\subsubsection{Target Model} Here, we consider an extended target model, where a sensory target (e.g. a human being or a vehicle) is located near to the BS, such that the sensing waveforms are reflected by various target components (scatters). Therefore, in a rich scattering environment, $\mathbf{R}_\text{g}$ is typically of full rank. Recalling that the transmitting and the receiving antennas employed for sensing are colocated at the BS,
thus, $\mathbf{R}_\text{g}$  can be modeled as \cite{9380222}
\begin{equation}
    \mathbf{R}_{\text{g}}= \sum_m \alpha_m^2 (\mathbf{a}(\theta_{m}) \mathbf{a}^H(\theta_{m})) \otimes (\mathbf{b}(\theta_{m}) \mathbf{b}^H(\theta_{m})), 
\end{equation}
where the scalar $\alpha_m$ is the radar cross section (RCS) of scatter $m$, $\mathbf{a}(\theta_{m})$ and $\mathbf{b}(\theta_{m})$ are the corresponding transmit and receive steering vectors to echo's direction of arrival $\theta_{m}$, respectively. Since we are only interested in the MSE-based waveform design at the transmitter side, the receive steering vectors are assumed to orthogonal with each other. Therefore, by introducing the transmit covariance matrix $\mathbf{R}_{\text{g}_\text{T}} = \sum_m \alpha_m^2 \mathbf{a}(\theta_{m}) \mathbf{a}^H(\theta_{m})$ and the receive covariance matrix $\mathbf{R}_{\text{g}_\text{R}} = \mathbf{I}_{\text{N}_\text{R}}$, we can simplify $\mathbf{R}_{\text{g}}$ as
\begin{equation}\label{R}
    \mathbf{R}_{\text{g}} = \mathbf{R}_{\text{g}_\text{T}} \otimes \mathbf{I}_{\text{N}_\text{R}}.
\end{equation}

%the target response matrix $\mathbf{G}$ can be written as,
%\begin{equation}
%    \mathbf{G} = \sum_m \alpha_m \mathbf{b}(\theta_{m}) \mathbf{a}^H(\theta_{m}),
%\end{equation}

\subsubsection{MSE-Based Transmit Waveform Design} From the sensing perspective, we consider an estimation-oriented waveform design that minimizes the MSE for target sensing. Since the received signal $\mathbf{y}_{\text{S}}$ and the target responses $\mathbf{g}$ are jointly Gaussian distributed, the MMSE estimator is indeed a linear MMSE (LMMSE) estimator. Recall the fact that the sensing waveform $\bar{\mathbf{X}}$ is known to the BS.
%The random vectors $\mathbf{y}_{\text{S}}$ and $\mathbf{g}$ are jointly Gaussian distributed as,
%\begin{equation}
%    \left[ \begin{array}{c}
%         \mathbf{g} \\
%         \mathbf{y}_{\text{S}}
%    \end{array} \right] \sim \mathcal{CN} \bigg(
 %     \left[ \begin{array}{c}
 %        \mathbf{0} \\
 %        \mathbf{0}
 %   \end{array} \right],
 %   \left[ \begin{array}{cc}
  %       \mathbf{R}_{\text{g}} & \mathbf{R}_{\text{g}}\bar{\mathbf{X}}^H  \\
 %        \bar{\mathbf{X}}\mathbf{R}_{\text{g}} & \bar{\mathbf{X}}\mathbf{R}_{\text{g}} \bar{\mathbf{X}}^H + \delta_{\text{S}}^2 \mathbf{I}
 %   \end{array} \right]
 %   \bigg). \nonumber
%\end{equation}
In particular, the MMSE estimator is expressed as
\begin{equation}
    \mathbf{g}_{\text{MMSE}} = (\bar{\mathbf{X}}^H\bar{\mathbf{X}} +\delta_{\text{S}}^2 \mathbf{I})^{-1}\bar{\mathbf{X}}^H\mathbf{y}_{\text{S}}.
\end{equation}
Therefore, the MSE matrix $\mathbf{E}$ is
\begin{align}
    \mathbf{E} = \mathbb{E}\{\|\mathbf{g} - \mathbf{g}_{\text{MMSE}}\|^2_2\}
   % & = \text{Tr} \{\mathbf{R}_{\text{g}}- \mathbf{R}_{\text{g}}\bar{\mathbf{X}}^H(\bar{\mathbf{X}}\mathbf{R}_{\text{g}} \bar{\mathbf{X}}^H + \delta_{\text{S}}^2 \mathbf{I})\bar{\mathbf{X}}\mathbf{R}_{\text{g}}\} \nonumber\\
     {=} \text{Tr}\{(\delta_{\text{S}}^{-2} \bar{\mathbf{X}}^H\bar{\mathbf{X}} + \mathbf{R}_{\text{g}}^{-1})^{-1}\}.
\end{align}
%where $(a)$ comes from the matrix inverse identity\footnote{ $(\mathbf{A}+\mathbf{BCD})^{-1}=\mathbf{A}^{-1}-\mathbf{A}^{-1}\mathbf{B}(\mathbf{DA}^{-1}\mathbf{B}+\mathbf{C}^{-1})^{-1}\mathbf{DA}$}. 
To bridge the connection between the sensing waveform design and communication precoder, we further observe that 
%\begin{equation} \label{XtoP}
    $\bar{\mathbf{X}}^H\bar{\mathbf{X}}=(\mathbf{X}\mathbf{X}^H) \otimes \mathbf{I}_{\text{N}_\text{R}} = \mathbf{R}_{\text{P}} \otimes \mathbf{I}_{\text{N}_\text{R}}.$
%\end{equation}
By taking \eqref{R} into account, the to-be-designed precoding matrix $\mathbf{P}$ appears in the MMSE sensing waveform design criteria. Exploiting the identity $\text{Tr}(\mathbf{A} \otimes \mathbf{I}_{\text{N}_\text{R}})= N_R\text{Tr}(\mathbf{A})$, the MMSE transmit waveform design problem is given by
\begin{equation}
\begin{split}
    \mathbf{E}(\mathbf{R}_{\text{P}}) & = \text{Tr}\{((\delta_{\text{S}}^{-2}\mathbf{R}_{\text{P}}+\mathbf{R}_{\text{g}_\text{T}}^{-1}) \otimes \mathbf{I}_{\text{N}_\text{R}})^{-1} \} \\
    &= N_R\text{Tr}\{(\delta_{\text{S}}^{-2} \mathbf{R}_{\text{P}}+ \mathbf{R}_{\text{g}_\text{T}}^{-1})^{-1}\}. 
\end{split}
\end{equation}
Therefore, when the transmitted communication signal is re-used for sensing, the waveform design problem that minimizes the target estimation error under the maximum power budget can be formulated as
\begin{align}
    \underset{\mathbf{R}_{\text{P}}}{\text{minimize}} \quad & \text{Tr}\{(\delta_{\text{S}}^{-2} \mathbf{R}_{\text{P}}+ \mathbf{R}_{\text{g}_\text{T}}^{-1})^{-1}\} \nonumber\\
    \text{s.t} \quad & \text{Tr}\{\mathbf{R}_{\text{P}}\} \leq P_T,  \mathbf{R}_{\text{P}} \succeq \mathbf{0}. \label{constraint11}
\end{align}
Obviously, the optimization problem \eqref{constraint11} is convex w.r.t $\mathbf{R}_{\text{P}}$. Its optimal point can be found by the well-known Karush–Kuhn–Tucker (KKT) conditions \cite{boyd2004convex}. % Note that in \eqref{constraint11}, communication performance is not considered. %That is, communication transmission may be even not available when the scatters and communication receivers are far apart from each other. Therefore, \eqref{constraint11} is not a practical ISAC waveform design strategy in most cases. 
Yet, in the next section, our analysis on the optimal solutions of (10) will unveil the parameter estimation performance of the downlink waveform in the monostatic ISAC system.

\section{A MSE Lower Bound} 
In this section, we analyze the parameter estimation performance by examining the above MSE problem. Here, we start with a MSE lower bound inspired from \eqref{constraint11}.

\textit{Theorem 1}: Let $\mathbf{R}_{\text{P}}$ be the covariance matrix of the BS precoder and $\mathbf{R}_{\text{g}_\text{T}}$ be the target response matrix, which are both $N_T \times N_T$ positive-definite full-rank Hermitian matrices. Let $\sigma_{i,\text{P}}$ and $\sigma_{i,\text{g}_\text{T}}$ be the $i$-th large eigenvalue of $\mathbf{R}_{\text{P}}$ and $\mathbf{R}_{\text{g}_\text{T}}$, respectively. Then, the MSE is lower-bounded by
\begin{equation}\label{MSEbound}
     \mathbf{E}(\mathbf{R}_{\text{P}}) \geq \sum_{i=1}^{N_T} (\delta_{\text{S}}^{-2}\sigma_{i,{\text{P}} } + \sigma_{i,\text{g}_\text{T}}^{-1})^{-1}.
\end{equation}
The equality holds if and only if the eigenvector matrix $\mathbf{U}_{\text{P}}$ = $\mathbf{U}_{\text{g}_\text{T}}$, where $\mathbf{U}_{\text{P}}$ and $\mathbf{U}_{\text{g}_\text{T}}$ consist of eigenvectors corresponding to descending ordered eigenvalues of $\mathbf{R}_{\text{P}}$ and $\mathbf{R}_{\text{g}_\text{T}}$, respectively.
\begin{proof}
It can be proved by following the similar procedure as in \cite{6086773} and is omitted here due to the page limitation.
\end{proof}

%$\{\sigma_{i,\text{P}}|\sigma_{i,\text{P}} \in \text{eig}(\mathbf{R}_{\text{P}})\}$ and $\{\sigma_{i,\text{g}_\text{T}}|\sigma_{i,\text{g}_\text{T}} \in \text{eig}(\mathbf{R}_{\text{g}_\text{T}})\}$
%$\mathbf{R}_{\text{P}} = \mathbf{U}_{\text{P}} \mathbf{\Lambda}_{\text{P}}^{\downarrow} \mathbf{U}_{\text{P}}^H$ and $\mathbf{R}_{\text{g}_\text{T}}= \mathbf{U}_{\text{g}_\text{T}} \mathbf{\Lambda}_{\text{g}_\text{T}}^{\downarrow} \mathbf{U}_{\text{g}_\text{T}}^H$ be the eigenvalue decomposition of $\mathbf{R}_{\text{P}}$ and $\mathbf{R}_{\text{g}_\text{T}}$, respectively, where $\mathbf{\Lambda}_{\text{P}}^{\downarrow} = \text{diag}\{\sigma_{1,\text{P}},\cdots,\sigma_{N_T,\text{P}}\}$ and $\mathbf{\Lambda}_{\text{g}_\text{T}}^{\downarrow} = \text{diag}\{\sigma_{1,\text{g}_\text{T}},\cdots,\sigma_{N_T,\text{g}_\text{T}}\}$ denote the corresponding eigenvalues organized in the ascending order. 

The MSE matrix evaluates the parameter estimation performance of sensing functionality. Typically, if $\mathbf{R}_{\text{g}_\text{T}}$ is known, \textit{Theorem 1} can be employed to find the optimal eigenvalues $\{{\sigma}^*_{i,{\text{P}}}\}_{i=1}^{N_T}$ that minimize the MSE. In \eqref{constraint11}, with the maximum power budget $P_
T$ at BS, the optimal eigenvalues is the water-filling solution given by
\begin{equation}
    \sigma^*_{i,{\text{P}}} = \left[\delta_{\text{S}}(\frac{1}{\sqrt{\lambda}}-\frac{\delta_{\text{S}}}{\sigma_{i,\text{g}_\text{T}}})\right]^+, \quad \sum_{i=1}^{N_T}\sigma^*_{i,{\text{P}}} = P_T.
\end{equation}
    Consequently, the optimal precoder can be designed as $\mathbf{P}=\mathbf{U}_{\text{P}}\text{diag}\{{\sigma^*}_{i,{\text{P}}}^{\frac{1}{2}},\cdots,{\sigma^*}_{N_T,{\text{P}}}^{\frac{1}{2}}\}$. If the to-be-designed $\mathbf{P}$ is unconstrained, it is obvious that the optimal solution is determined by $\mathbf{R}_{\text{g}_\text{T}}$'s eigenvalues and the corresponding eigenvectors. However, in a practical downlink monostatic ISAC sensing system, the transmitted waveform matrix is often rank-deficient.

Typically, in the communication BS, the precoder $\mathbf{P}$ delivers $K$ data streams, stacking as $\mathbf{S}$, to its corresponding users via $N_T$ transmitting antennas. If we re-use the communication waveform $\mathbf{X}$ for sensing, the rank of the transmitted sensing waveform is given by
\begin{equation}\label{rankequation}
    \text{rank}(\mathbf{R}_{\text{X}}) \overset{\eqref{X2P}}{=} \text{rank}(\mathbf{R}_{\text{P}}) \leq \min\{K,N_T\}.
\end{equation}

By examining \eqref{MSEbound} and \eqref{rankequation}, we further observe that \eqref{MSEbound} still holds when $K = N_T$. However, when $K < N_T$, there are $N_T-K$ eigenvalues $\{{\sigma}_{i,{\text{P}}}\}_{i=K+1}^{N_T}$ of $\mathbf{R}_{\text{P}}$ that are zero, inspiring the following corollary.

\textit{Corollary 1}: Let $\mathbf{R}_{\text{P}}$ be the rank-deficient covariance matrix of the BS precoder, where $\text{rank}(\mathbf{R}_{\text{P}}) = K$ and $K < N_T$. Then, the MSE is given by
\begin{equation}\label{MSEbound2}
     \mathbf{E}(\mathbf{R}_{\text{P}})  \geq \sum_{k=1}^{K}(\delta_{\text{S}}^{-2}\sigma_{k,{\text{P}} } + \sigma_{k,\text{g}_\text{T}}^{-1})^{-1} +\sum_{i=K+1}^{N_T} \sigma_{i,\text{g}_\text{T}}.
\end{equation}
The equality holds if and only if the eigenvector matrix $\mathbf{U}_{\text{P}}$ = $\mathbf{U}_{\text{g}_\text{T}}$, where $\mathbf{U}_{\text{P}}$ and $\mathbf{U}_{\text{g}_\text{T}}$ consist of eigenvectors corresponding to descending ordered eigenvalues of $\mathbf{R}_{\text{P}}$ and $\mathbf{R}_{\text{g}_\text{T}}$, respectively. 

\begin{proof} The proof is given as
\begin{equation}\nonumber
\begin{split}
     \mathbf{E}(\mathbf{R}_{\text{P}})  =& \text{Tr}\{(\delta_{\text{S}}^{-2} \mathbf{R}_{\text{P}}+ \mathbf{R}_{\text{g}_\text{T}}^{-1})^{-1}\}  \geq  \sum_{i=1}^{N_T} (\delta_{\text{S}}^{-2}\sigma_{i,{\text{P}} } + \sigma_{i,\text{g}_\text{T}}^{-1})^{-1}\\
    =& \sum_{k=1}^{K}(\delta_{\text{S}}^{-2}\sigma_{k,{\text{P}} } + \sigma_{k,\text{g}_\text{T}}^{-1})^{-1} +\sum_{i=K+1}^{N_T} (\delta_{\text{S}}^{-2}\sigma_{i,{\text{P}} } + \sigma_{i,\text{g}_\text{T}}^{-1})^{-1}\\
   =& \sum_{k=1}^{K}(\delta_{\text{S}}^{-2}\sigma_{k,{\text{P}} } + \sigma_{k,\text{g}_\text{T}}^{-1})^{-1} +\sum_{i=K+1}^{N_T} \sigma_{i,\text{g}_\text{T}}.
\end{split}
\end{equation}
\end{proof}
\textit{Corollary 1} indicates that in the downlink monostatic ISAC system, if the transmitted communication waveform is re-used for sensing, the MSE matrix is governed by the rank of the transmitted waveform, which is indeed, controlled by the number of served communication users. Moreover, the estimation performance attains its MSE bound when the user number $K$ is equal to the antenna numbers $N_T$.
On the other hand, when the transmitted signal power is unlimited, which implies that all $\sigma_{i,{\text{P}}}$ of $\mathbf{R}_{\text{P}}$ are sufficiently large, \eqref{MSEbound} and \eqref{MSEbound2} can then be characterized as
\begin{equation}
     \liminf_{\{\sigma_{i,{\text{P}}}\}_{i=1}^{N_T}\rightarrow \infty} \mathbf{E}(\mathbf{R}_{\text{P}}) = 0,
\end{equation}
and
\begin{equation}
     \liminf_{\{\sigma_{i,{\text{P}}}\}_{i=1}^{K}\rightarrow \infty} \mathbf{E}(\mathbf{R}_{\text{P}}) =  \sum_{i=K+1}^{N_T} \sigma_{i,\text{g}_\text{T}},
\end{equation}
respectively, it can be seen that the $\mathbf{E}(\mathbf{R}_{\text{P}})$ is lower bounded by $0$ when $\mathbf{R}_{\text{X}}$ is full rank, however, $\mathbf{E}(\mathbf{R}_{\text{P}})$ will always be positive because the eigenvalues of the target echoes $\sum_{i=K+1}^{N_T}\sigma_{i,\text{g}_\text{T}}$ are always positive. In other words, the sensing MSE of the downlink monostatic ISAC system could not always achieve the lower bound $0$ for $K < N_T$, if the transmitted communication waveform is re-used without any further processing.

%Therefore, the gap between the MSE lower bound and the ISAC sensing performance can be measured by,
%\begin{equation}
 %   \text{Gap}\overset{\eqref{MSEbound}-\eqref{MSEbound2}}{=} \sum_{i=K+1}^{N_T} \frac{\delta_{\text{S}}^{-2}\sigma_{i,{\text{P}}}\sigma_{i,\text{g}_\text{T}}}{\delta_{\text{S}}^{-2}\sigma_{i,{\text{P}} } + \sigma_{i,\text{g}_\text{T}}^{-1}}
%\end{equation}

\section{DoF completion : A Scheme to Achieve MSE Lower Bound} 
In this section, we introduce a waveform augmentation strategy to improve the sensing performance of the downlink monostatic ISAC system, namely DoF completion. The main idea is to complete the transmitted signal from a rank-deficient matrix to a full rank one, by embedding an additional signal structure into the transmitted communication signals. Therefore, the resultant waveform dedicated for sensing contains not only the re-used communication signal, but also an augmented waveform matrix to extend the spatial DoFs. To this end, we express the augmented waveform $\hat{\mathbf{X}}$ as
\begin{equation}
    \hat{\mathbf{X}} = [\mathbf{P} \quad \mathbf{P}_{\text{A}}] \begin{bmatrix}
    \mathbf{S}\\
    \mathbf{S}_{\text{A}}
    \end{bmatrix} \in \mathbb{C}^{N_T \times L},
\end{equation}
where $\mathbf{P}_{\text{A}}$ and $\mathbf{S}_{\text{A}}$ are the $N_T \times (N_T -K)$ and $(N_T -K) \times L$ additional sensing precoding matrix and data matrix, respectively. We assume that the additional data matrix is a unitary matrix such that
$
    \mathbf{S}_{\text{A}}\mathbf{S}^H_{\text{A}} = \mathbf{I}_{N_T-K}.
$
Therefore, the covariance matrix of $\hat{\mathbf{X}}$ can be written as
\begin{equation}
      \mathbf{R}_{\hat{\text{X}}} = [\mathbf{P} \quad  \mathbf{P}_{\text{A}}] \begin{bmatrix}
    \mathbf{P}^H\\
    \mathbf{P}_{\text{A}}^H
    \end{bmatrix}   = \mathbf{R}_{\text{P}}+ \mathbf{R}_{\text{P}_{\text{A}}} =  \mathbf{R}_{\hat{\text{P}}},
\end{equation}
where $\mathbf{R}_{\text{P}_{\text{A}}} \in \mathbb{C}^{N_T \times N_T}$ is the covariance matrix dedicated to the additional precoder $\mathbf{P}_{\text{A}}$ and $\mathbf{R}_{\hat{\text{P}}}$ denotes the covariance matrix of the augmented precoder matrix. 

Hence, when the augmented signal waveform is employed, the resulting MSE matrix that measures the sensing performance can then be written as
\begin{equation}
     \mathbf{E}(\mathbf{R}_{\hat{\text{P}}}) =  N_R\text{Tr}\{(\delta_{\text{S}}^{-2} \mathbf{R}_{\hat{\text{P}}}+ \mathbf{R}_{\text{g}_\text{T}}^{-1})^{-1}\}. 
\end{equation}
It is also worth noting that the to-be-designed matrix is now  $\mathbf{R}_{\hat{\text{P}}}$ instead of the previsou $\mathbf{R}_{\text{P}}$ for DoF completion. Recalling \textit{Corollary 1}, the MSE matrix $\mathbf{E}(\mathbf{R}_{\hat{\text{P}}})$ is also lower bounded by the eigenvalues of $\mathbf{R}_{\hat{\text{P}}}$ and $\mathbf{R}_{\text{g}_\text{T}}^{-1}$ as
\begin{equation}\label{MSEbound3}
     \mathbf{E}(\mathbf{R}_{\hat{\text{P}}}) \geq \sum_{i=1}^{N_T} (\delta_{\text{S}}^{-2}\sigma_{i,{\hat{\text{P}}} } + \sigma_{i,\text{g}_\text{T}}^{-1})^{-1}.
\end{equation}
Consequently, the proposed strategy is able to achieve the MSE lower bound of 0 when the transmitted signal power is unlimited, i.e.,
\begin{equation}
     \liminf_{\{\sigma_{i,{\hat{\text{P}}}}\}_{i=1}^{N_T}\rightarrow \infty} \mathbf{E}(\mathbf{R}_{\hat{\text{P}}}) = 0.
\end{equation}

\textbf{Remark}: In general, the idea behind DoF completion  strategy is to generate a transmit waveform that is full-rank. In the radar context, a full-rank waveform matrix is able to guarantee the feasibility of unbiased estimation \cite{liu2021cram}, as well as to realize omni-directional beampattern without additional mechanical rotation systems or phased-array beam scanning. In the above analysis, we proved that the full-rank waveform could achieve the MSE lower bound of 0 for a sufficient large transmit power, which implies that the parameter estimation errors from target echoes can be minimized. However, the above strategy still brings several drawbacks. On the one hand, the emission of the additional signal structure will always take additional costs of the transmission power. Intuitively, these additional power is dedicated for sensing but may not necessarily be beneficial for the communication functionality. On the other hand, while the additional signal structure from the DoF completion strategy can be removed by the successive interference cancellation (SIC) technology, this requires high computation capability at the communication users' side. Therefore, we introduce the following MMSE waveform design approach to deal with the 
dual-functional waveform design problem.

\section{MMSE Waveform Designs via DoF completion } 

In this section, we apply the DoF completion strategy into a monostatic downlink ISAC waveform design problem. Our objective function is to minimize the MSE of the sensing performance, under the per-user SINR constraint for communication functionality. Note that this is a typical waveform design problem in the monostatic downlink ISAC scenario. Therefore, the waveform design can be formulated as
\begin{align}
    \underset{\hat{\text{P}}}{\text{minimize}}\quad & \text{Tr}\{(\delta_{\text{S}}^{-2} \mathbf{R}_{\hat{\text{P}}}+ \mathbf{R}_{\text{g}_\text{T}}^{-1})^{-1}\} \label{MMSEwaveform}\\
   \text{s.t} \quad  &  \gamma_k(\text{P}) \geq \gamma_0, \forall k \in \mathcal{K},\label{constraint1}\\
    & \text{Tr}\{\mathbf{R}_{\hat{\text{P}}}\} \leq P_T,   \mathbf{R}_{\hat{\text{P}}} \succeq \mathbf{0}.\label{constraint2}
\end{align}
where the received SINR of each communication user is required to be larger than the threshold $\gamma_0$ given by \eqref{constraint1} and \eqref{constraint2} limits the overall transmit power at the BS. When \eqref{constraint1} is removed or $\gamma_0 = 0$, the waveform design concerns only the target estimation performance that is equivalent to \eqref{constraint11}. % On the other hand, the problem in \eqref{constraint1} and \eqref{constraint2} may be infeasible if $\gamma_0$ is not properly chosen.
Due to the SINR constraint \eqref{constraint1} for communication functionality, the overall problem is non-convex. Fortunately, \eqref{constraint1} can be reformulated by employing SDR technology, $\forall k \in \mathcal{K}$,
\begin{equation}
    \mathbf{Q}_k =\mathbf{p}_k\mathbf{p}_k^H \Longleftrightarrow \mathbf{Q}_k \succeq \mathbf{0}, \text{rank}(\mathbf{Q}_k) = 1,
\end{equation}
and $\mathbf{R}_{\text{P}} = \sum_{k\in \mathcal{K}}\mathbf{Q}_k$. We then recast constraint \eqref{constraint1} as
\begin{equation}
    \text{tr}(\mathbf{H}_k\mathbf{Q}_k)\geq \gamma_0 (\sum_{j \in \mathcal{K}/k} \text{tr}(\mathbf{H}_k\mathbf{Q}_k) + \delta_C^2), \forall k, \text{rank}(\mathbf{Q}_k) = 1,
\end{equation}
where $\mathbf{H}_k = \mathbf{h}_k\mathbf{h}_k^H $. Similarly, by defining $\{\mathbf{Q}_k\}_{k = K+1}^{N_T} = \sum_{k = K+1}^{N_T}\mathbf{p}_k\mathbf{p}_k^H$, we could find $\mathbf{R}_{\text{P}_\text{A}} = \sum_{k = K+1}^{N_T}\mathbf{Q}_k$. Then, the optimization problem can be reformulated as
\begin{equation}
\begin{aligned}
     \underset{\{\mathbf{Q}_k\}_{k = 1}^{N_T}}{\text{minimize}}
\quad & \text{Tr}\{(\delta_{\text{S}}^{-2}  \sum_{k = 1}^{N_T} \mathbf{Q}_k+ \mathbf{R}_{\text{g}_\text{T}}^{-1})^{-1}\} \\
   \text{s.t} \quad  &   \frac{\text{tr}(\mathbf{H}_k\mathbf{Q}_k)}{\sum_{j \in \mathcal{K}/k} \text{tr}(\mathbf{H}_k\mathbf{Q}_k) + \delta_C^2} \geq \gamma_0, \forall k\in \mathcal{K}\\
    & \text{Tr}\{\sum_{k = 1}^{N_T}\mathbf{Q}_k\} \leq P_T,  \text{rank}(\mathbf{Q}_k) = 1. \label{constraint3}
\end{aligned}    
\end{equation}
By dropping the rank-one constraint $\text{rank}(\mathbf{Q}_k) = 1$, i.e., the relaxed problem is a convex one that can be solved by the well-known CVX toolbox. However, in this case, the resulting optimal solution is a relaxed solution of \eqref{constraint3}, which is typically not guaranteed to yield a rank-one optimal point. By employing Theorem 4 in \cite{liu2021cram}, we could construct a tight rank-one solution. Denoting the relaxed optimal solution as $\{\Hat{\mathbf{Q}}_k\}_{k = 1}^{N_T}$ with convariance matrix $\Hat{\mathbf{R}}_{\hat{\text{P}}}$ , the rank-one beamformers $\mathbf{p}_k$ in $\mathbf{P}$ can be constructed by
\begin{equation}\label{optP}
    \mathbf{p}_k = (\mathbf{h}_k^H\Hat{\mathbf{Q}}_k\mathbf{h}_k)^{-1/2}\Hat{\mathbf{Q}}_k\mathbf{h}_k, \forall k \in \mathcal{K},
\end{equation}
and $\mathbf{P}_{A}$ can always be obtained from
\begin{equation}\label{optPA}
    \mathbf{P}_{\text{A}}\mathbf{P}_{\text{A}}^H = \Hat{\mathbf{R}}_{\hat{\text{P}}} - \sum_{k \in \mathcal{K}}\Hat{\mathbf{Q}}_k.
\end{equation}
Note \eqref{optP} and \eqref{optPA} yield an optimal solution of \eqref{MMSEwaveform}. Therefore, the above MMSE waveform design problem is solved optimally.
%One way to find an acceptable rank-one approximation solution is to construct a matrix using the eigenvector $\mathbf{u}_{1,\text{Q}_k^*}$ corresponding the largest eigenvalue $\delta_{1,\text{Q}_k^*}$, where $\mathbf{Q}_k^*$ is the optimal solution of \eqref{constraint3}. Then, the rank-one approximated solutions $\hat{\mathbf{Q}}_k^*$ can be constructed as, $\forall k$,
%\begin{equation}
 %  \hat{\mathbf{Q}}_k^* = \delta_{1,\text{Q}_k}\mathbf{u}_{1,\text{Q}_k^*}\mathbf{u}_{1,\text{Q}_k^*}^H.
%\end{equation}
%By defining $\mathbf{U}_{\hat{\text{P}}} = \{\mathbf{u}_{1,\text{Q}_1^*},\cdots,\mathbf{u}_{1,\text{Q}_{N_T}^*}\}$, where $\mathbf{u}_{1,\text{Q}_1^*}$ is the eigenvector corresponding the largest eigenvalue of optimal $\mathbf{Q}_1^*$, then the optimal precoder can be given as,
%\begin{equation}
 %   \hat{\mathbf{P}}^* = \mathbf{U}_{\hat{\text{P}}}\text{diag}\{{\sigma^*}_{1,{\text{Q}_1^*}}^{\frac{1}{2}},\cdots,{\sigma^*}_{1,{\text{Q}_{N_T}^*}}^{\frac{1}{2}}\}%.
%\end{equation}
%Before going futher

\section{Simulations}
In this section, we provide numerical results to verify the advantages of our proposed DoF completion strategy, as well as to show the performance of the MSE-based sensing waveform design. Without loss of generality, the ISAC BS is equipped with $N_R = 10$ receive antennas, under the maximun power budget of $P_T = 40$ dBm. The transmission frame length $L = 30$, the carrier frequency is set as $2.4$ GHz, the noise powers are $\delta_{\text{C}}^2 = \delta_{\text{S}}^2 = -100$ dBm \cite{wei2019performance}. The communication channel experiences Rayleigh fading, where each entry of the channel matrix follows the standard complex Gaussian distribution. The path loss exponent is $2.2$ according to the 3GPP path loss model. Besides, $K$ users are
randomly and uniformly distributed in a circle centered at (40 m, 0 m) with a radius of 10 m. Moreover, we assume that the entries of the target response matrix $\mathbf{G}$ follows Swerling 2 model with Gaussian distributed complex amplitude \cite{liu2021cram,richards2014fundamentals}.

\begin{figure}[!t]
	\centering 
	\includegraphics[width=0.8\columnwidth]{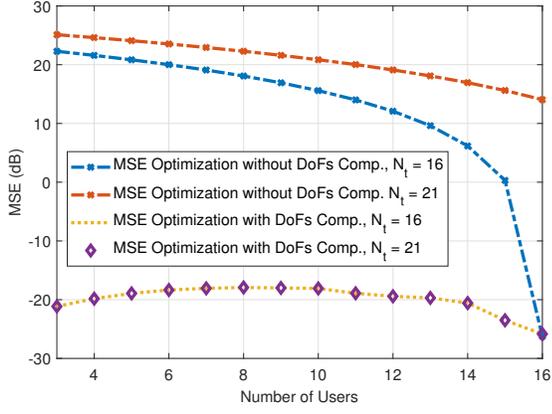}
	\caption{The sensing estimation performances versus the number of communication users with and without our proposed DoF completion  strategy at $N_T = 16$ and $N_T = 21$.}
	\label{Fig2}
\end{figure}

In Fig.~\ref{Fig2}, we compare the estimation performance of the sensing functionality in terms of MSE versus with the growth of $K$ with and without our proposed DoF completion strategy. The number of antennas are set to be $N_T = 16$ and $N_T = 21$, respectively. Without DoF completion strategy, the increase of $K$ leads to a lower MSE. This observation is due to the fact that the rank of the transmit waveform grows with the number of users that facilitates the design to achieve the MSE lower bound in \eqref{MSEbound2}. On the contrary, for the schemes without DoF completion, the resulting MSE performance is not affected by the number of users, which is indeed the MSE lower bound. More interestingly, when $K = N_T = 16$, the MSE performance without DoF completion also achieves the lower bound. This observation validates the correctness of our MSE analysis in Section IV, where the performance gap between the MSE lower bound and performance of optimal MSE solution vanishes at $K = N_T$.

\begin{figure}[!t]
	\centering 
	\includegraphics[width=0.8\columnwidth]{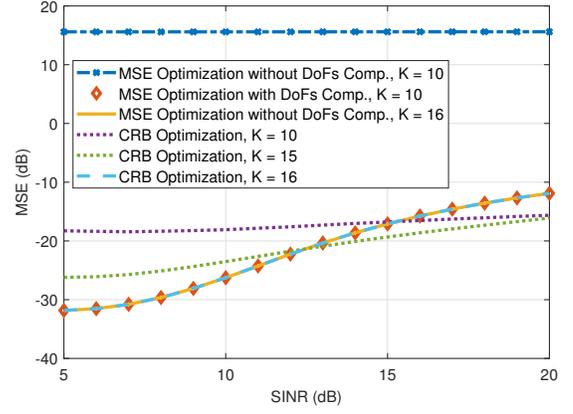}
	\caption{The sensing estimation performances versus the minimum required SINR to the communication users with and without the proposed DoF completion  strategy.}
	\label{Fig3}
\end{figure}

In Fig.~\ref{Fig3}, we evaluate the performance of SINR constrained MMSE waveform design for the downlink monostatic ISAC system \eqref{constraint3}. The Cramer-Rao Bound (CRB) based waveform design in \cite{liu2021cram} acts as a baseline. We observe that without the DoF completion strategy, the MMSE waveform design can achieve the CRB at $K= N_T = 16$. More interestingly, by employing the propsoed DoF completion strategy, the performance of MMSE waveform design is no longer affected by the number of served communication users. The comparison between the MMSE waveform designs with and without DoF completion strategy shows the performance gain of the proposed strategy.

%Here we consider a statistic sensing model where each element of $\mathbf{G}$ is random variable with independent and identical distribution.

\section{Conclusion}
In this letter, the estimation performance of downlink monostatic ISAC system was investigated. We derived a parameter estimation lower bound that facilitated the derivation of the downlink monostatic ISAC waveform, and then proposed a waveform augmentation strategy to improve the parameter estimation performance. The proposed strategy was adopted in the MMSE ISAC waveform design constrained by the per-user SINR of communication functionality. The MSE lower bound and the effectiveness of the proposed waveform design were revealed by the simulations.


\bibliographystyle{IEEEtran}
\bibliography{IEEEabrv,bare_conf}


%\vfill


\end{document}